\documentclass[sigconf,10pt,nonacm]{acmart}
\setcopyright{none}

\usepackage{wrapfig}
\usepackage{comment}
\usepackage{colortbl}

\usepackage{hyperref}

\usepackage{amsthm} %

\usepackage{color,soul}
\definecolor{lightblue}{rgb}{.4,.8,1}
\sethlcolor{lightblue}
\usepackage{graphics}
\usepackage{hyperref}
\usepackage{lipsum}
\usepackage{tikz}
\usepackage{enumitem}	%
\usepackage{listings} %
\usepackage{booktabs}	%
\usepackage{lipsum}

\usepackage{capt-of}	%
\usepackage{diagbox}
\usepackage{multirow}
\usepackage{todonotes}  %

\definecolor{refkey}{rgb}{249,158,26}
\definecolor{labelkey}{rgb}{0,1,0}
\definecolor{airforceblue}{rgb}{0.36, 0.54, 0.66}
\definecolor{applegreen}{rgb}{0.55, 0.71, 0.0}

\usepackage{mathrsfs}	%
\usepackage{amsfonts}

\usepackage[export]{adjustbox} %

\usepackage[plain]{algorithm2e}
\usepackage{boldline}					%

\usepackage{multirow}

\definecolor{frenzyorange}{RGB}{249, 158, 26}
\newcommand*\circled[1]{\tikz[baseline=(char.base)]{
			\node[shape=circle,draw,text=black,inner sep=1pt] (char) {#1};}}

\renewcommand{\paragraph}[1]{\vskip 3pt\noindent\textbf{#1 }}	 %

  {\begin{list}{$\bullet$}%
     {\setlength{\parsep}{0pt}%
      \setlength{\topsep}{0pt}%
      \setlength{\itemsep}{2pt}}}%
  {\end{list}}
\newcommand\Noted[1]{} %

\definecolor{darkblue}{rgb}{0.0, 0.0, 0.55}
\definecolor{mygreen}{HTML}{ADFF2F}
\definecolor{mylightgray}{gray}{0.8}

  {\begin{itemize}
	[leftmargin=0cm,
		itemindent=.3cm,
		labelwidth=\itemindent,
		labelsep=0pt,
		parsep=1pt,
		topsep=1pt,
		itemsep=1pt,
		align=left]
  }%
  {\end{itemize}}    

  {\begin{enumerate}
	[leftmargin=.cm,itemindent=.5cm,labelwidth=\itemindent,
		labelsep=0pt,
		parsep=1pt,
		topsep=1pt,
		itemsep=3pt,
		align=left]
  }%
  {\end{enumerate}}

\newcommand{\sys}{SpQt}
\newcommand{\stepnum}[1]{\circled{\scriptsize #1}}  %

\makeatletter
\def\@copyrightspace{\relax}
\makeatother

\begin{abstract}
Large language models (LLMs) increasingly run on personal devices, where generating text is slow: producing each
token requires reading all of the model's weights from memory. Two techniques reduce this memory traffic. Quantization
stores the weights in 4 to 8 bits. Activation sparsity treats near-zero activations as zero and skips the weights
they multiply; prior work shows that it skips 40--50\% of the weights of modern LLMs without retraining, at little
loss in accuracy, also on quantized models. Yet the two do not combine in execution: a quantization format packs the
weights that multiply many consecutive activations into one block, and a block can be skipped only if all of these
activations are near zero, which rarely happens. Moreover, the near-zero activations change with every token, so
balancing the GPU's load and finding them add run-time overhead.
We present \sys{}, which makes activation sparsity practical for quantized LLMs. \sys{} (1) stores each activation's
quantized weights contiguously, in a \emph{zigzag layout} that keeps the quantization formats unchanged; (2) divides the
nonzero activations evenly among groups of GPU threads, so the load stays balanced; and (3) finds
the nonzero activations inside kernels that each decode step already runs before its matrix products, such as
normalization, so that finding them adds no kernel launch. Built into llama.cpp and tested on five devices from a phone to a desktop, with Apple and NVIDIA
GPUs and four quantization formats, \sys{} generates tokens 1.45--2.15$\times$ faster than llama.cpp at 50\%
sparsity, with up to 2.6$\times$ less energy per token. With \sys{}, a device can choose a sparsity level in addition to
a quantization format, trading a small loss in accuracy for speed and energy.
\end{abstract}

\begin{document}

\title{Practical activation sparsity for quantized LLMs on personal devices}

\author{Rongxiang Wang}
\authornote{This work was started during an internship at Zoom.}
\affiliation{%
	\institution{University of Virginia}
	\city{Charlottesville}
	\state{Virginia}
	\country{USA}
}
\email{waq9hw@virginia.edu}
\author{Kangyuan Shu}
\affiliation{%
	\institution{Shopify}
	\city{Bellevue}
	\state{Washington}
	\country{USA}
}
\email{kangyuan.shu@shopify.com}
\author{Felix Xiaozhu Lin}
\affiliation{%
    \institution{University of Virginia}
	\city{Charlottesville}
	\state{Virginia}
	\country{USA}
}
\email{felixlin@virginia.edu}

\date{}

\thispagestyle{empty}
\maketitle
\section{Introduction}
\label{sec:intro}

\begin{figure}[t]
\includegraphics[width=\columnwidth]{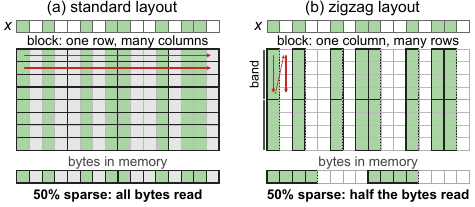}
\caption{In a group-wise quantized layout, skipping the zeroed activations saves almost no memory traffic (a); the
zigzag layout stores each column's quantized weights contiguously, so skipping a zeroed activation skips its whole
column (b). Green: nonzero activations and their active columns; gray: read but not needed.}
\label{fig:layout}
\end{figure}

Large language models (LLMs) increasingly run on personal devices such as phones, laptops, small desktops and
embedded boards, for privacy, availability and cost~\cite{apple-intelligence, autodroid, laskaridis2024melting}. On these devices
generating text is bound by memory: each decode step multiplies every weight matrix by a single vector,
so every token reads every weight once, while the arithmetic is cheap. Quantization is therefore the
norm: engines such as llama.cpp~\cite{ggerganov2022llamacpp} store weights in 4- to 8-bit formats that cut the bytes
per token by 2--4$\times$ against 16-bit weights~\cite{Kawrakow2023kquant, frantar2022gptq}.

\paragraph{Activation sparsity} Activation sparsity promises a second, independent saving. A projection multiplies its
weight matrix by an input vector, whose entries are the \emph{activations}: activation $x_j$ multiplies column $j$ of
the weights. In a modern LLM, many activations are close to zero, and zeroing those below a per-projection magnitude
threshold leaves the output nearly unchanged~\cite{liu2025teal, lee2024cats}. This works on models with non-ReLU
activation functions as they are, with no retraining or learned predictor, which the neuron sparsity of earlier systems
requires~\cite{liu2023dejavu, song2024powerinfer}. TEAL reaches 40--50\% sparsity on Llama-2, Llama-3 and
Mistral models with little loss in accuracy, also on quantized weights~\cite{liu2025teal}. A decode step then needs
only the columns of the nonzero activations, the \emph{active columns}, and could read half the weights at 50\%
sparsity. TEAL, however, left the kernels that would skip the other columns in quantized weights to future work.

\paragraph{Challenges} Yet the two techniques do not combine in execution, for three reasons. First, group-wise
formats pack the weights that multiply many consecutive activations into one block (\autoref{fig:layout}a), and a
block can be skipped only when all of these activations are zeroed: at 50\% sparsity, such blocks hold 0.2--2\% of
the weight bytes. Skipping the zeroed activations one by one inside a block still reads the whole block and
adds branches, which slows a GPU kernel down (\S\ref{sec:motiv-conflict}). Second, the active columns fall unevenly: a GPU kernel
that gives each threadgroup a fixed range of columns, as PowerInfer's does~\cite{song2024powerinfer}, suffers load
imbalance among its threadgroups, more so when the active columns cluster. Third, the nonzero activations depend
on the token being decoded, so they are known only at run time and each decode step must find them. Finding them in separate GPU kernels would add kernel
launches to a decode step that already runs many small kernels, such as normalizations and residual additions.

\paragraph{Our approach} We present \sys{}, a system that makes activation sparsity practical for quantized LLMs on
personal devices. Our key insight is that each challenge can be removed by redesigning one part of the decode step:
the quantized layout can store each activation's weights contiguously, keeping its quantization format, so that skipping a
zeroed activation skips a chunk of memory and its compute; the GPU kernel can be co-designed with the new layout to divide the active
columns evenly among threadgroups; and the decode step's compute graph can be rearranged so that kernels it already
runs, such as normalization, also find the nonzero activations, with no kernel launches of their own. \sys{} realizes this
insight with three techniques.

\emph{(1) The zigzag layout.} Within a band of rows, \sys{} quantizes each block as 256 rows of \emph{one column}
and stores a column's blocks contiguously (\autoref{fig:layout}b). The weights of a zeroed activation are then
contiguous in memory, so they can be skipped as whole blocks, while the formats stay llama.cpp's Q4\_K, Q5\_K, Q6\_K and Q8\_0.

\emph{(2) The zigzag sparse GEMV kernel.} Every GPU threadgroup takes an equal slice of an \emph{active-column list},
the sorted indices of a projection's active columns, and adds its sums into an output zeroed in advance, so no threadgroup waits for another.

\emph{(3) Fused sparsity detection.} For every input vector of a layer, one kernel launch writes the
active-column list of each projection that reads it, without atomics. \sys{}
builds the lists inside the normalization, SiLU and residual kernels, so finding the active columns adds no kernel
launch.

\paragraph{Results} We implement \sys{} in llama.cpp for Apple GPUs (Metal) and NVIDIA GPUs (CUDA), and test it on five
devices: an iPhone Air, a MacBook Pro (M4 Pro), a Mac Studio (M2 Max), a Jetson Orin Nano and a Jetson AGX Xavier, with
Llama-2 7B/13B/70B, Llama-3-8B and Mistral-7B in four formats. At 50\% sparsity, \sys{} decodes tokens
1.45--2.15$\times$ faster than llama.cpp, with 1.8--2.6$\times$ less energy per token on the Macs and the Jetsons. The
cost of sparsity in accuracy is nearly the same in every format, as TEAL found for
other quantization methods~\cite{liu2025teal}, so format and sparsity can be traded against each other. For instance,
Llama-3-8B in Q5\_K at 25\% sparsity is as accurate as the popular Q4\_K\_M preset and decodes faster on every device.

\paragraph{Contributions} This paper makes three contributions.
\begin{itemize}[leftmargin=*, nosep]
\item We identify why activation sparsity, accurate on quantized models, fails to speed them up on GPUs, and measure
  the causes on real activations.
\item We present \sys{}, which co-designs the quantized layout, the GPU kernel and the decode step's compute graph: the
  zigzag layout, which keeps llama.cpp's formats while making a zeroed activation's weights skippable; the zigzag
  sparse GEMV kernel; and fused sparsity detection.
\item We implement \sys{} in llama.cpp for Metal and CUDA and evaluate its speed, accuracy and energy on five devices.
\end{itemize}
\section{Motivations}
\label{sec:motiv}

\subsection{Decoding on personal devices}

\paragraph{A decode step} A decoder layer of a Llama-family model has seven \emph{linear projections}, each of which
multiplies an input vector by a weight matrix: the query, key, value and output projections of attention ($W_q, W_k,
W_v, W_o$) and the gate, up and down projections of the feed-forward network ($W_g, W_u, W_d$). When the model
generates text, each decode step processes one token, so each projection is a matrix-vector product (GEMV) that reads
every weight once and performs two operations per
weight. The GPUs of personal devices offer 68--400\,GB/s of memory bandwidth, so the step is
bound by reading the weights. In our measurements on the M4 Pro, these weight products take 87--92\% of a decode step.

\paragraph{A primer on GPU execution} The CPU records the kernels of a decode step in a command buffer (a CUDA stream
or graph on NVIDIA GPUs), and the GPU runs them in order. Each kernel \emph{launch} runs as a grid of \emph{threadgroups}
(thread blocks in CUDA). Each threadgroup runs on one GPU core, where its threads share a fast \emph{threadgroup memory}, and its threads execute in
\emph{SIMD groups} of 32 \emph{lanes} (warps in CUDA) that run in lockstep and can exchange values by
shuffles. Threadgroups run independently and combine results only through device memory, for instance with atomic
additions. A dense GEMV therefore gives each threadgroup its own output rows, whose dot products it computes alone.
A decode step is a sequence of launches: per layer, the seven GEMVs and thirteen small kernels, such as
normalizations, attention and residual additions. Each launch that depends on the previous one waits for it to finish
and then refills the GPU, so even a launch with little work costs time.

\paragraph{Quantization formats} Quantization reduces the bytes that the GEMVs read. The K-quants of
llama.cpp~\cite{Kawrakow2023kquant}, widely used on personal devices, divide each row of a weight matrix
into \emph{superblocks} of 256 consecutive weights, each stored with its own scales: 4.5 bits per weight in Q4\_K, 5.5
in Q5\_K and 6.6 in Q6\_K; Q8\_0 stores blocks of 32 8-bit values and a scale. GPTQ, AWQ and other group-wise schemes
likewise quantize each row in groups of consecutive weights, for example 128~\cite{frantar2022gptq, lin2024awq}. The
blocks follow the GEMV: each output is the dot product of one row with the input vector, so a kernel walks each row
block by block, dequantizes a block, and multiplies it with the activations of its columns.

\subsection{Activation sparsity}

For a projection $y = Wx$, activation sparsity zeroes the activations $x_j$ whose magnitude is below a threshold $t$:
$y \approx \sum_{j : |x_j| > t} x_j W_{:,j}$, so only the active columns of $W$, those of the nonzero activations,
are needed. TEAL showed that this works without the ReLU function~\cite{liu2025teal}. In modern LLMs, whose
feed-forward networks use the SiLU function, few activations are exactly zero, but the activations at every projection
are concentrated around zero, and a threshold per projection, calibrated once on about 20,000 tokens of text, reaches
40--50\% sparsity with little loss in accuracy.
TEAL's greedy search allocates sparsity across the projections for a model-level target, the \emph{effective
sparsity}: the share of a layer's weights that a decode step skips. The allocation differs by projection. In
Llama-3-8B at a 50\% target, $W_v$ skips 36\% of its columns on average and $W_o$ 62\%; $W_q$, $W_k$ and $W_v$ share
one input vector but skip at three different rates.

A related kind, neuron sparsity, skips the feed-forward neurons that a learned predictor expects a token to leave
inactive, as in DejaVu and PowerInfer~\cite{liu2023dejavu, song2024powerinfer}. It skips the most on models with the
ReLU function, which modern models regain only through retraining~\cite{mirzadeh2024relu, song2024prosparse}.
Activation sparsity needs neither predictors nor retraining.

The ideal speedup of activation sparsity is simple: at sparsity $s$ a GEMV reads a fraction $1-s$ of the weights, so
50\% sparsity could halve the time of the weight products. Its accuracy also carries over to quantized weights: TEAL
found that its errors and those of 8-bit RTN, 4-bit AWQ and 2--3-bit QuIP\# quantization compound roughly
independently, but left the kernels that would combine the two to future work~\cite{liu2025teal}. The rest of this
section shows why such kernels are hard to build.

\subsection{Activation sparsity conflicts with quantization}
\label{sec:motiv-conflict}
\begin{table}[t]
\caption{The share of a layer's weight bytes that a decode step could skip at the 50\% target, measured on WikiText-2:
in row blocks of $B$ columns, and in the zigzag layout.}
\label{tab:skip}
\footnotesize
\setlength{\tabcolsep}{2pt}
\begin{tabular*}{0.9\columnwidth}{@{\extracolsep{\fill}}lccc@{}}
\toprule
 & \textbf{Llama-2-7B} & \textbf{Llama-3-8B} & \textbf{Mistral-7B} \\ \midrule
\multicolumn{4}{@{}l}{Row blocks of $B$ columns:} \\
\quad 256 (K-quants) & 0.28\% & 0.18\% & 0.20\% \\
\quad 128 (GPTQ, AWQ) & 0.97\% & 0.32\% & 0.36\% \\
\quad 32 (Q8\_0) & 1.95\% & 1.40\% & 1.13\% \\
\quad 4 & 9.8\% & 8.7\% & 8.8\% \\ \addlinespace
Zigzag layout & 49.1\% & 48.8\% & 48.8\% \\
\bottomrule
\end{tabular*}
\end{table}

The two techniques group the weights along different axes. Sparsity selects columns of $W$; quantization groups
weights along rows. A Q4\_K superblock covers 256 columns, so a zeroed activation accounts for 1/256 of each block that
contains its column, and a block can be skipped only when all 256 of its activations are zeroed. To see how often
this happens, we decoded WikiText-2 with TEAL's thresholds and counted at every step the blocks of $B$
consecutive columns whose activations are all zeroed (\autoref{tab:skip}). In six of the seven projections the active
columns are spread as if at random: their runs of consecutive active columns are as long as random ones, to within
0.1\%. Only the input of $W_o$, where the outputs of the attention heads lie side by side, clusters: at 50\%, 14--22\%
of its 32-column blocks are empty, against 0.9--1.4\% at random. Over a whole layer, row blocks of 256 columns could
skip 0.2--0.3\% of the weight bytes at the 50\% target and even blocks of 4 under 10\%, while the zeroed activations
leave out 49\%. Almost every block is read at every sparsity a model can tolerate (\autoref{fig:layout}a).

Skipping inside a block saves at most the multiply-adds of the zeroed activations, while the loads remain, and each
skip adds a branch to the kernel's inner loop. We built this baseline, \emph{Stock-Skip}, by adding
the skip to llama.cpp's own GEMV kernels. At 50\% sparsity it runs at 0.17--0.72$\times$ the speed of the original
kernel on three Apple GPUs. On two NVIDIA GPUs it runs at 0.96--0.98$\times$: there llama.cpp's kernel converts the
activations to 8-bit integers and multiplies four of them by four weights in one instruction, so it can skip only
groups of four zeroed activations, which hold under 10\% of the weights (\autoref{tab:skip}). On the standard layout, skipping slows the
GEMV down on every GPU we test; \S\ref{sec:eval-gain} will compare Stock-Skip with \sys{}.

\paragraph{How about other layouts and formats?} First, \emph{unquantized weights stored by column}, as in TEAL's
kernels, let a zeroed activation skip its column but give up quantization: at 50\% sparsity, 16-bit weights move more
bytes than a 4.5-bit format read densely. Second, \emph{fewer bits} save bytes too, but cost more accuracy than
sparsity does for the same time; \S\ref{sec:eval-sens} will compare the two.

\paragraph{Design implications} Quantizing the weights along the columns, so that each block holds consecutive rows of
one column, keeps both the skip and the format; PowerInfer's transposed down projection is one
instance~\cite{song2024powerinfer}. Such a layout still leaves three problems. First, it calls for a new split of
the GEMV's work: a dense GEMV gives each threadgroup its own output rows, and with blocks along the columns
the work must be divided over both the rows and the active columns. Second, the active columns change every token, and their
number in a fixed range of columns varies from range to range, more so at the input of $W_o$, where they cluster; a
fixed split of the columns thus causes load imbalance among the threadgroups. Third, the active columns
depend on the token being decoded and are known only at run time, so each decode step must find them for every
projection; separate kernels for this would add launches to a step that already runs many small ones, such as
normalizations and residual additions. These problems motivate three goals for the layout and the GPU kernels:
(i) the layout stores a zeroed activation's weights contiguously, so that they can be skipped as a whole, and keeps the
quantization formats, and hence their accuracy, unchanged; (ii) the GPU's load stays balanced as the active columns
change; and (iii) the nonzero activations are found every step without extra kernel launches.
\section{Overview}
\label{sec:overview}

\paragraph{System model} We target personal devices with a GPU (phones, laptops, desktops and embedded boards) that
run an LLM in a group-wise quantized format and decode one user's requests one token at a time. \sys{} is training-free and
runs entirely on the device. It adds three parts to the on-device inference stack: a quantizer that converts a
16-bit model into the zigzag layout; Metal and CUDA kernels behind two new operators, one
that computes a sparse GEMV on weights in the zigzag layout and one that builds the active-column lists; and a compute
graph that uses the operators. As in TEAL, sparsity applies only to the seven projections in every layer, which multiply weights by activations.

\begin{figure}[t]
\includegraphics[width=\columnwidth]{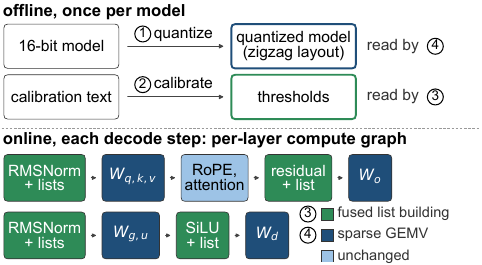}
\caption{The architecture and workflow of \sys{}: the offline steps (top) and the per-layer compute graph of a decode
step (bottom).}
\label{fig:overview}
\end{figure}

\paragraph{The operation} \autoref{fig:overview} shows how \sys{} runs. Offline, once per model, a quantizer converts
the 16-bit weights into a quantized model of the chosen format in the zigzag layout (\stepnum{1}; \S\ref{sec:design-layout}), and a calibration run over about 20,000 tokens of text sets each
projection's threshold for each target sparsity, with TEAL's allocation across the projections (\stepnum{2};
\S\ref{sec:impl}). At load time, \sys{} reads the quantized model and the thresholds of the chosen sparsity. A prompt is processed densely, as one batch,
as in TEAL, and the first token of a sequence is never sparsified. Every later decode step takes the sparse path. In
each layer, before each projection, a fused operator performs one of the step's small operations, such as the RMS
normalization, and in the same kernel builds the projection's active-column list and initializes its output (\stepnum{3};
\S\ref{sec:design-lists}). The projection's sparse GEMV then splits that list evenly over the GPU's threadgroups and
reads and computes only the active columns (\stepnum{4};
\S\ref{sec:design-gemv}).

\paragraph{Applicability} The layout applies to group-wise formats in general, because it keeps each format's block
encoding and changes only which weights a block holds. The lists and the GEMV serve any projection
with many near-zero activations, the experts of mixture-of-experts models included. They run on both Apple and NVIDIA
GPUs, because they use only primitives that both GPU families offer (\S\ref{sec:design-gpus}).

\paragraph{Limitations} First, \sys{} targets decoding one token at a time. A batch of tokens needs the union of their
active columns, which approaches all columns as the batch grows, so batches, such as those of speculative decoding,
are future work. Second, prompts are computed densely, so requests dominated by a long prompt and a short answer gain
little (\S\ref{sec:eval-requests}). Third, \sys{} runs on GPUs only: the NPUs of phones and laptops execute dense,
statically scheduled tensor programs, which do not readily support skipping a different set of columns every token.

\section{Design}
\label{sec:design}

\subsection{The zigzag layout}
\label{sec:design-layout}

The layout must store a zeroed activation's weights contiguously without changing the format. Consider a K-quant,
which quantizes 256 consecutive weights as one superblock with its scales. \sys{} sets the \emph{unit} to 256
consecutive weights of one column and quantizes each unit as one superblock. It then cuts the rows of each weight matrix into \emph{bands} of $H = 256n$ rows ($n = 1, 2, 4$),
so that each column of a band holds $n$ units. A band stores each column's $n$ units next to each other, in column
order, and the bands follow each other in row order (\autoref{fig:layout}b).

Skipping a zeroed activation now skips its column's units, which are contiguous in every band, so at sparsity $s$ a
sparse GEMV reads a fraction $1-s$ of the weight bytes. The quantization format keeps its bits per weight, block
structure and dequantization, so a model file in the zigzag layout is as large as one in the original layout. Its
quantization error differs slightly: in our measurements its perplexity is at most 0.3\% above the original layout's
(\S\ref{sec:eval-accuracy}). The bands split the work by rows: a band's partial sums fit in one threadgroup
(\S\ref{sec:design-gemv}), and storing a band's units together keeps a threadgroup's reads within one region of memory.

\paragraph{Choosing the band} The band height trades off three costs. (1) For each active column, every band first
reads the column's index and its activation, then decodes the column's $n$ units, so a taller band spreads these extra
reads over more weight bytes. (2) A taller band leaves fewer bands to divide among the GPU's threadgroups. (3) A taller band needs more
threadgroup memory to combine its sums. The best band thus depends on the GPU, and we use the fastest band of each
device (\S\ref{sec:impl}).

\subsection{The zigzag sparse GEMV kernel}
\label{sec:design-gemv}

\begin{figure}[t]
\includegraphics[width=\columnwidth]{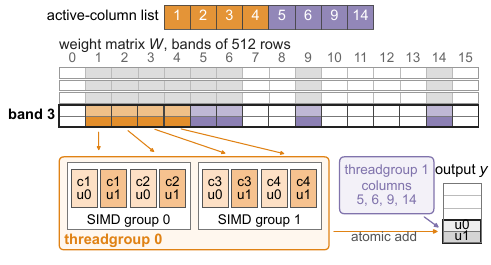}
\caption{The zigzag sparse GEMV kernel on one band of a weight matrix. c1 u0: unit 0 of column 1.}
\label{fig:gpu}
\end{figure}

The zigzag layout and activation sparsity call for a new GEMV kernel. Stock llama.cpp's kernels give each SIMD group
one or a few rows of the weight matrix, and the SIMD group reads all of their blocks. On the zigzag layout, the kernel
computes only the active columns of each band. These columns differ from token to token and are unevenly distributed
across the band. The kernel must therefore distribute them evenly among threadgroups, regardless of where the active columns
lie. To this end, the kernel takes the projection's active-column list as input, built before the kernel runs
(\S\ref{sec:design-lists}). It splits the work two ways
(\autoref{fig:gpu}): the bands split the rows, and the list of $N$ active columns is cut into $S$ equal slices. It runs
a grid of (bands $\times$ splits) threadgroups, and threadgroup $(b, k)$ computes the slice
$[\lfloor kN/S \rfloor, \lfloor (k+1)N/S \rfloor)$ of the list in band $b$.
In \autoref{fig:gpu}, band 3 has eight active columns and two threadgroups, so each threadgroup computes four active
columns, although six of the eight lie in the first half of the band. Every threadgroup thus gets the same amount of
work. By contrast, giving each threadgroup
a fixed range of columns, as \emph{Zigzag-Fixed} does, causes load imbalance: the ranges hold different numbers of
active columns, some full and others empty when activity clusters (\S\ref{sec:eval-gain}).

\paragraph{Mapping a slice to lanes} Within a threadgroup, the SIMD groups interleave through the slice; in
\autoref{fig:gpu}, threadgroup 0's two SIMD groups take two of its four columns each. As in llama.cpp's
Metal kernels for Q4\_K and Q5\_K~\cite{ggerganov2022llamacpp}, a SIMD group of 32 lanes forms four 8-lane groups,
each of which decodes one 256-weight block, 32 weights per lane; \sys{} uses this mapping for every format. Here the
block is a unit, so a lane's 32 weights lie in 32 rows
of one column: the lane dequantizes them, multiplies them by the column's activation, and keeps 32 running sums, one
per row, in registers. With $n$
units per column, the $n$ units of a column go to $n$ neighboring lane groups, so a SIMD group takes $4/n$ columns
per step and every lane always works on an active column. In \autoref{fig:gpu}, $n = 2$: unit u0 of a column holds the
band's upper 256 rows and u1 the lower 256, and each unit's sums go to those rows of the output $y$. To hide memory latency, each lane group keeps $U$
units \emph{in flight}: it reads the next $U$ entries of its slice and starts loading all $U$ units before it decodes
the first, so that the loads overlap. Only the dequantization depends on the format: each format has its own unit
decoder, which follows the format's dequantization in llama.cpp, and the rest of the kernel is shared.

\paragraph{Combining the sums} At the end, lane groups that hold the same rows combine their sums by SIMD
shuffles, and the SIMD groups combine theirs through threadgroup memory. Each threadgroup then adds its band's sums
into the output with float atomics; the output is zeroed, or initialized with the residual, when the lists are built
(\S\ref{sec:design-lists}).

\subsection{Fused sparsity detection}
\label{sec:design-lists}

\begin{figure}[t]
\includegraphics[width=\columnwidth]{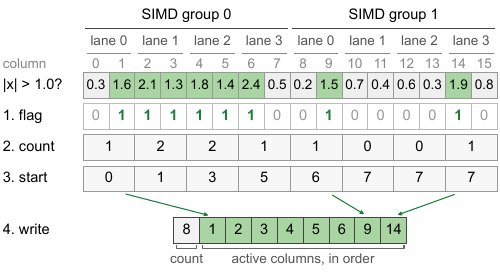}
\caption{Building a sorted active-column list from flags, counts and prefix sums, without atomics.}
\label{fig:lists}
\end{figure}

Every decode step must give the GEMV each projection's active-column list and an initialized output, without adding kernel launches to the step. A layer has four distinct input vectors for its projections: the normalized
input of $W_q$, $W_k$ and $W_v$; the attention output, input of $W_o$; the normalized input of $W_g$ and $W_u$; and
$\mathrm{SiLU}(W_g x) \odot W_u x$, input of $W_d$. Each projection has its own threshold, so one input vector yields
up to three lists, one per projection that reads it (\S\ref{sec:motiv}).

\paragraph{Building a list} Building a list is stream compaction, a standard GPU
primitive~\cite{sengupta2007scan, billeter2009compaction}. To build the lists, one threadgroup of up to 1,024 threads scans the input vector
(\autoref{fig:lists}).
Each thread takes a few consecutive entries, sets one bit of a flag word for each entry above each list's
threshold, and counts its flagged entries per list. Prefix sums of these counts, computed within each SIMD group by
shuffles and then across SIMD groups through threadgroup memory, give each thread the position of its first flagged
entry in each list, where it writes the indices of its flagged entries. The lists come out sorted, as
\texttt{[count, j$_1$, j$_2$, \dots]}. With sorted lists, the
GEMV reads a band's units in increasing address order.

\paragraph{Fusing list building into existing kernels} We build the lists in the kernels that already compute
the input vectors they index. Each of these kernels first computes its input vector and then builds the
lists from it (\autoref{fig:overview}). For the inputs of $W_{q,k,v}$ and $W_{g,u}$, the input vector is the RMS
normalization multiplied by its learned weights; for the input of $W_d$, it is $\mathrm{SiLU}(g) \odot u$. Meanwhile
the kernel's other threadgroups initialize the output array of the projection's sparse GEMV
(\S\ref{sec:design-gemv}): they zero it, or, for $W_o$ and $W_d$, copy the residual stream into it. Because the GEMV
accumulates its results into this array, it also produces the residual sum $x + W a$. The lists, the output array
and the computed input vector share one buffer, which the GEMV reads and accumulates into. We also stack $W_q$,
$W_k$, $W_v$ and $W_g$, $W_u$ into one matrix each, computed by one GEMV whose bands read the list of their own
projection, and apply the rotary position embedding (RoPE) to the queries and keys in a single kernel launch.

In a decode step of Llama-3-8B (Q4\_K) on the M4 Pro at 50\%, with each kernel launch timed on its own, a layer's
eight small kernel launches for the normalizations and their weights, SiLU and residual additions become four kernel
launches that also build the lists. The four take less GPU time than the eight did (28.3 against 31.8\,$\mu$s, medians
of 3 runs). Finding the active columns thus adds no kernel launch.

\subsection{Portability across GPUs}
\label{sec:design-gpus}

\begin{table}[t]
\caption{Launch rules per GPU family. (Groups = SIMD groups per threadgroup; Units = units in flight per lane group;
Target = SIMD groups per kernel launch, Q4\_K / Q5\_K / Q6\_K / Q8\_0.)}
\label{tab:rules}
\scriptsize
\setlength{\tabcolsep}{1.5pt}
\begin{tabular*}{0.9\columnwidth}{@{\extracolsep{\fill}}lccc@{}}
\toprule
\textbf{GPU family} & \textbf{Groups} & \textbf{Units} & \textbf{Target} \\ \midrule
Apple8 (M2 Max) & 4 & 1--2 & 512/512/384/384 \\
Apple9/10 (M4 Pro, A19 Pro) & 2 & 1 & 384/384/768/768 \\
Ampere (Orin Nano) & 8--16 & 1--2 & 384/256/384/384 \\
Volta (AGX Xavier) & 4--8 & 1--2 & 1024/1024/384/1024 \\
\bottomrule
\end{tabular*}
\end{table}

The design maps onto the GPUs of both vendors. SIMD groups and threadgroups on Apple GPUs (Metal) correspond
to warps and thread blocks on NVIDIA GPUs (CUDA), and both GPU families provide the SIMD shuffles, threadgroup memory
and floating-point atomic operations that our kernels use.

\paragraph{Launch rules} Three launch parameters of the zigzag sparse GEMV kernel (\S\ref{sec:design-gemv}) depend on the format, the weight matrix and the GPU: (1) the
number of SIMD groups per threadgroup, (2) the number of units each lane group keeps in flight, and (3) the number of
splits per band. One rule per GPU family sets them (\autoref{tab:rules}). The number of splits is the power of two
that brings a kernel launch closest to a target number of SIMD groups; the other parameters are fitted by sweeping
the kernel over the matrix shapes of our models at several sparsity levels. The rules differ across families: on the M2 Max and the M4 Pro, the other
Apple family's rule slows the Q4\_K GEMV by 13--16\%. On Apple9/10 GPUs, the larger units of Q6\_K and Q8\_0 need
more SIMD groups per kernel launch, and the Xavier's Volta GPU prefers smaller thread blocks than the Orin Nano's
Ampere GPU.

\paragraph{CUDA graphs} On the Orin Nano, the decode step runs as a CUDA graph, as in llama.cpp, and is captured again
when the thresholds change after the dense first token.
      
\section{Implementation}
\label{sec:impl}

We implement \sys{} in 8.8K SLoC of C++, Metal and CUDA atop llama.cpp~\cite{ggerganov2022llamacpp} (release b5711):
the Metal kernels (3.6K), the CUDA kernels (1.4K), two new operators in ggml, llama.cpp's tensor library (0.8K), the
compute graph and the quantizer (1.6K), and tools (1.4K). Tests and benchmarks take another 5K.

\paragraph{Operators and kernels} The design's GPU kernels sit behind the two operators. One computes a projection on
weights in the zigzag layout: the zigzag sparse GEMV kernel (\S\ref{sec:design-gemv}) for one token, and a dense product
for a prompt. The other builds the active-column lists and initializes the GEMV's output (\S\ref{sec:design-lists}); it
can first compute the input vector itself, by a normalization or the SiLU function. Each kernel is written once in
Metal, for macOS and iOS, and once in CUDA. A compute graph for Llama-architecture models uses the two operators in
place of llama.cpp's projections; attention and the output layer remain llama.cpp's.

\paragraph{Quantizer} Our quantizer is built on llama.cpp's. It converts a model's 16-bit GGUF file into a given format
and band: it stacks $W_{q,k,v}$ and $W_{g,u}$, reorders the weights of each matrix into the zigzag layout
(\S\ref{sec:design-layout}), and encodes them with llama.cpp's own routines, unchanged. As in llama.cpp, it works from
the trained weights alone, with no retraining, and quantizes every weight matrix of the decoder layers to the chosen
format. Conversion is a one-time step per model and device, in the device's fastest band: 1,024 rows on the iPhone Air, 512 on
the Macs and 256 on the Jetsons.

\paragraph{Calibration} We adopt TEAL's calibration method~\cite{liu2025teal} and reimplement it in llama.cpp, so that
the activations are measured on the quantized model that a device runs. A calibration tool runs the model over TEAL's
calibration text, ten sequences of 2,048 tokens from its Alpaca samples~\cite{taori2023alpaca}, and records, per layer,
histograms of the activation magnitudes in the four input vectors. A script then takes the per-projection sparsity
allocation of TEAL's greedy search and reads each projection's threshold from the histograms; it writes one threshold
file per model file and target sparsity. The same tool measures the sparsity that a threshold file reaches in decoding,
with TEAL's measure, which weights each projection by its size.

\paragraph{The dense path} At load time, \sys{} reads the threshold file of the chosen sparsity. Decode steps take the
sparse path. Prompts and the first token of a sequence are computed densely (\S\ref{sec:overview}), on the same weights
in the zigzag layout. On
Apple GPUs, the dense path is llama.cpp's simdgroup-matrix kernel, adapted to read units in the zigzag layout. On NVIDIA
GPUs, it dequantizes whole bands to half precision in chunks and multiplies them with cuBLAS.

\paragraph{Evaluation mode} Perplexity and multiple-choice tasks score a given text in one forward pass and decode no
token, so \sys{} would compute them densely, as a prompt. To measure sparse decoding on them, we follow
TEAL~\cite{liu2025teal}: the first half of each sequence is computed densely and the second half sparsely, as if it were
decoded. An evaluation mode computes the sparse half in batches: it zeroes the activations below each projection's
threshold, which equals skipping their columns. A generation task runs as deployed: a dense prompt, then sparse decode
steps.
\providecommand{\pending}[1]{\textcolor{red}{[#1]}}

\section{Evaluation}
\label{sec:eval}

We answer the following questions:
\begin{itemize}[leftmargin=*, nosep]
\item How much faster does \sys{} decode, at what cost in accuracy and with how much energy (\S\ref{sec:eval-e2e})?
\item How much does each key design contribute (\S\ref{sec:eval-gain})?
\item How do the benefits change with the request, the context and the calibration text (\S\ref{sec:eval-sens})?
\end{itemize}

\begin{figure*}[t]
\includegraphics[width=\textwidth]{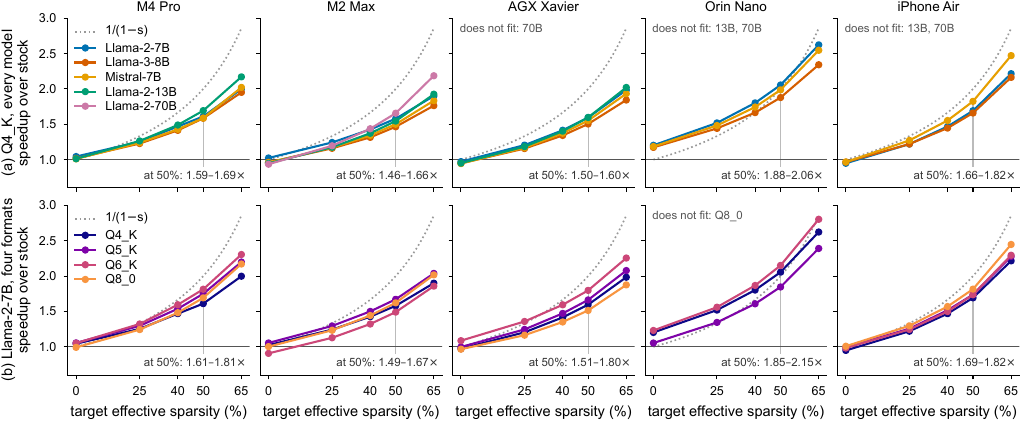}
\caption{\sys{} decodes 1.45--2.15$\times$ faster than stock llama.cpp at 50\% sparsity, on every device: decode-step
speedup for (a) every model that fits a device, in Q4\_K, and (b) Llama-2-7B in four formats.}
\label{fig:headline}
\end{figure*}

\subsection{Methodology}
\label{sec:eval-setup}

\begin{table}[t]
\caption{The five devices. GPU: architecture and compute units (Apple GPU cores, NVIDIA SMs).}
\label{tab:devices}
\footnotesize
\setlength{\tabcolsep}{1.5pt}
\begin{tabular*}{0.9\columnwidth}{@{\extracolsep{\fill}}llrr@{}}
\toprule
\textbf{Device} & \textbf{GPU} & \textbf{Mem.} & \textbf{GB/s} \\ \midrule
MacBook Pro M4 Pro & Apple9, 20 cores & 48\,GB & 273 \\
Mac Studio M2 Max & Apple8, 30 cores & 64\,GB & 400 \\
Jetson AGX Xavier & Volta, 8 SMs & 32\,GB & 102 \\
Jetson Orin Nano & Ampere, 8 SMs & 8\,GB & 68 \\
iPhone Air (A19 Pro) & Apple10, 5 cores & 12\,GB & $\sim$76 \\
\bottomrule

\end{tabular*}
\end{table}

\paragraph{Test platforms} We test on five devices (\autoref{tab:devices}): a phone (iPhone Air, A19 Pro), a laptop
(MacBook Pro, M4 Pro), a desktop (Mac Studio, M2 Max) and two embedded boards (NVIDIA's Jetson Orin Nano and Jetson AGX
Xavier, held at their 15\,W and 30\,W power modes). Their GPUs come from two vendors and offer 68--400\,GB/s of memory
bandwidth.

\paragraph{Models and formats} We run Llama-2-7B and 13B~\cite{touvron2023llama2}, Llama-3-8B~\cite{dubey2024llama},
Mistral-7B~\cite{jiang2023mistral} and Llama-2-70B in llama.cpp's Q4\_K, Q5\_K, Q6\_K and Q8\_0
formats~\cite{ggerganov2022llamacpp, Kawrakow2023kquant}, each on every device whose memory holds it.

\paragraph{Sparsity} We evaluate target effective sparsities of 25, 40, 50 and 65\%, with thresholds calibrated as in
\S\ref{sec:impl}. We also run \sys{} at 0\%, without thresholds, to measure what
it costs without sparsity.

\paragraph{Metrics} We follow the common practice.

\underline{Speed.} We report the time of a decode step, end to end with the output layer, over 128 decoded tokens after
a 128-token prompt: the mean of 3 rounds of 8 sequences, or on the iPhone the median of 8 runs, each started 60\,s after
iOS reports its coolest thermal state (nominal).

\underline{Accuracy.} We follow TEAL's protocol and measure on the quantized models, in \sys{}'s evaluation mode
(\S\ref{sec:impl}). We report (1) WikiText-2 perplexity~\cite{merity2017pointer} over 128 windows of 2,560 tokens; and
(2) the accuracy on TEAL's six tasks, scored by the evaluation harness~\cite{eval-harness} on a fixed subset of each
(300 to 1,140 questions): 5-shot MMLU~\cite{hendrycks2021mmlu}, 25-shot ARC-Challenge~\cite{clark2018arc}, 10-shot
HellaSwag~\cite{zellers2019hellaswag}, 5-shot GSM8K~\cite{cobbe2021gsm8k}, and zero-shot PIQA~\cite{bisk2020piqa} and
WinoGrande~\cite{sakaguchi2020winogrande}.

\underline{Energy.} We report the energy per decoded token, measured over the decode steps from the CPU, GPU and Neural
Engine power on the Macs (\texttt{powermetrics}) and from the power rails on the Jetsons, as the median over runs.

\paragraph{Baselines} We compare the following designs.
\begin{itemize}[leftmargin=*, nosep]
\item \emph{Stock}: llama.cpp at the version our fork builds on, with the same model, format and settings. It is our
main baseline.
\item \emph{Stock-Skip}: a kernel baseline (\S\ref{sec:eval-gain}). It adds a skip of the zeroed activations to
stock's GEMV kernels (\S\ref{sec:motiv-conflict}).
\item \emph{Zigzag-Fixed}: a kernel baseline (\S\ref{sec:eval-gain}). It runs on the zigzag layout but gives each
threadgroup a fixed range of columns and tests every activation, as PowerInfer's down projection
does~\cite{song2024powerinfer}.
\item \emph{PowerInfer}~\cite{song2024powerinfer}: a system built on llama.cpp that skips the feed-forward neurons
its learned predictors expect to be inactive. We compare end to end, on its own ReLU model.
\end{itemize}

\subsection{End-to-end results}
\label{sec:eval-e2e}

\begin{figure*}[t]
\includegraphics[width=\textwidth]{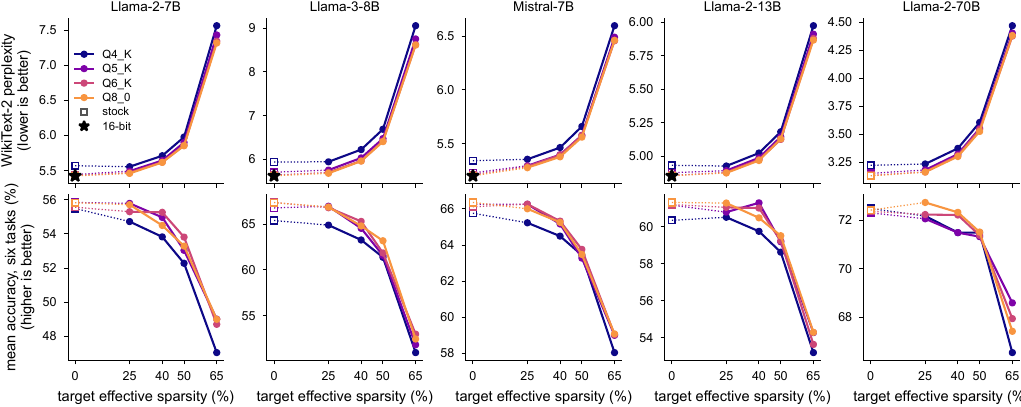}
\caption{Sparsity costs nearly the same accuracy in every format: WikiText-2 perplexity (top) and the mean accuracy of
six tasks (bottom) against effective sparsity.}
\label{fig:accuracy}
\end{figure*}

\paragraph{Speed} \sys{} speeds up the decode step on every device, model and format, and more so as sparsity grows
(\autoref{fig:headline}). At 50\% sparsity it decodes 1.45--2.15$\times$ faster than stock across all models and
formats; the gain grows from 1.12--1.56$\times$ at 25\% to 1.76--2.80$\times$ at 65\%. For Llama-3-8B in Q4\_K at
50\%, a decode step shrinks from 71.8 to 43.3\,ms on the iPhone Air and from 122.1 to 65.0\,ms on the Orin Nano. The
phone thus generates 23 tokens per second, up from 14.

At 50\%, the gain is the largest on the Orin Nano (1.72--2.15$\times$) and the smallest on the M2 Max and the Xavier
(1.45--1.84$\times$). Across models, the 70B gains the most (1.66$\times$ on the M2 Max in Q4\_K at 50\%) and
Llama-3-8B the least on every device. This is because the weight products take a larger share of a larger model's step
(\S\ref{sec:eval-gain}), while Llama-3-8B's output layer, over a vocabulary of 128k tokens, is fixed work. The gain
holds in all four formats (\autoref{fig:headline}b): on every device, the speedups of Llama-2-7B in the four formats lie
within 0.3$\times$ of each other at 50\%. The power mode changes the gain little: on the Xavier, Llama-3-8B in Q4\_K at
50\% gains 1.50--1.53$\times$ in the 15\,W, 30\,W and MAXN modes.

\paragraph{Accuracy}\label{sec:eval-accuracy} The accuracy cost is small up to 25\% sparsity and grows beyond it
(\autoref{fig:accuracy}). Up to 25\%, perplexity is at most 1.5\% above stock's, and
the six-task mean stays within 0.8 points of stock's. At 50\%, perplexity rises by 5.1--13.9\%, the least for
Llama-2-13B and the most for Llama-3-8B, and the six-task mean is 0.9--5.5 points lower. Math suffers the most: GSM8K
falls from 42.3 to 35.0 for Llama-3-8B in Q4\_K at 50\%. At 65\% the cost is large (perplexity
+21--53\%). The cost is nearly the same in every format: Llama-3-8B's perplexity at 50\% rises by 12.6\% in Q4\_K and by 13.9\% in
Q8\_0. The errors of sparsity and quantization thus add up roughly independently in llama.cpp's formats too, as TEAL
observed for other quantization methods~\cite{liu2025teal}, which lets a device trade one for the other
(\S\ref{sec:eval-sens}).

\paragraph{Cost without sparsity} Without sparsity, \sys{} changes speed and accuracy little. At 0\% sparsity, \sys{} runs
at 0.88--1.11$\times$ of stock's speed on the Apple GPUs and the Xavier. On the Orin Nano it runs at
1.03--1.23$\times$, because stock's CUDA kernel is slower there than ours even without sparsity. \sys{}'s perplexity
at 0\% is at most 0.3\% above stock's on the \mbox{7--13B} models (not shown), because the two layouts group the weights into
blocks in different ways.

\begin{figure}[t]
\includegraphics[width=\columnwidth]{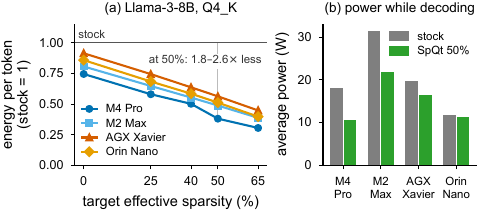}
\caption{\sys{} takes 1.8--2.6$\times$ less energy per token at 50\%, as both the time and the power drop: (a)
energy per decoded token relative to stock; (b) average power while decoding, stock against 50\%.}
\label{fig:energy}
\end{figure}

\paragraph{Energy} \sys{} reduces the energy per token (\autoref{fig:energy}): at 50\%, Llama-3-8B in Q4\_K takes
1.8--2.6$\times$ less energy per token on the Macs and the Jetsons. The energy drops because a decode step
takes less time and draws less power: at 50\%, the power drops from 18.1 to 10.6\,W on the M4 Pro and from 31.4 to
21.8\,W on the M2 Max. \sys{} draws less power than stock already at 0\% sparsity (24.3\,W on the M2 Max), and
less still as it skips weights. The exception is the Orin Nano, whose board draws about 11.5\,W either way, so its
energy follows its time.

\begin{figure}[t]
\includegraphics[width=\columnwidth]{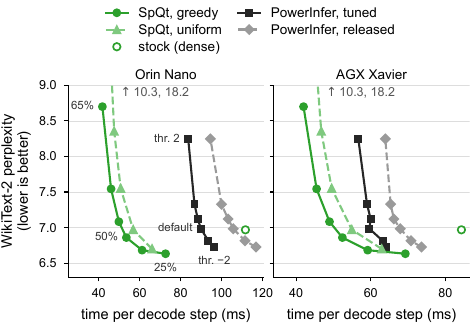}
\caption{\sys{} reaches each of PowerInfer's accuracies in less time: WikiText-2 perplexity against time per decode step
for ReluLLaMA-7B on the two Jetsons. \sys{} runs at 25, 40, 50, 55, 60 and 65\% sparsity and PowerInfer at predictor
thresholds (thr.) from \textminus 2 to 2 (default: 0, {\boldmath$\sim$}54\% sparsity).}
\label{fig:powerinfer}
\end{figure}

\paragraph{Comparison with PowerInfer}\label{sec:eval-powerinfer} \sys{} decodes faster than PowerInfer~\cite{song2024powerinfer} at every perplexity that PowerInfer reaches (\autoref{fig:powerinfer}).
PowerInfer's public code runs only ReLU models with trained predictors, so we compare them
on one of its models, ReluLLaMA-7B (Llama-2-7B fine-tuned to the ReLU function), on the two Jetsons, which hold
the whole model in GPU memory. PowerInfer runs its released model, quantized with its own tool to Q4\_0, the only quantized format its sparse kernels
support (4.5 bits per weight, same as Q4\_K), with every neuron on the GPU, at predictor thresholds from $-2$ to 2
(default: 0). We run it as released. PowerInfer was built for desktop GPUs, so we also tune its down-projection kernel
for each Jetson and run the \emph{tuned} version. \sys{} runs the model in Q4\_K from 25 to 65\% sparsity.
Its thresholds follow TEAL's greedy allocation. We also run it with uniform thresholds, which target the same sparsity
in every projection and need no search. PowerInfer builds on a llama.cpp version of December 2023, without flash attention;
\sys{} runs with flash attention on the Orin Nano and without it on the Xavier. Both systems run the workload and the
perplexity protocol of \S\ref{sec:eval-setup}, each in its own code (medians of at least three rounds).

At PowerInfer's default threshold (perplexity 6.98), \sys{} at 50\% is more accurate (6.86) and 1.68$\times$ faster than
tuned PowerInfer on the Orin Nano (53.5 against 89.8\,ms per step) and 1.14$\times$ faster on the Xavier (52.4 against
59.6\,ms). Against PowerInfer as released, it is 1.98 and 1.30$\times$ faster. The same holds at all of
PowerInfer's thresholds. \sys{} with uniform thresholds also outperforms PowerInfer.

PowerInfer gains less from the weights it skips. At its default threshold it skips 54\% of a layer's weight bytes, as
many as \sys{} at 55\%, yet \sys{} at 55\% runs 1.80 and 1.22$\times$ faster than tuned PowerInfer. This is because
PowerInfer's predictors add overhead, and its down projection splits the neurons into fixed
ranges without balancing, as \emph{Zigzag-Fixed} does (\S\ref{sec:eval-gain}); the tuned version decodes 14--18\% faster
but keeps this scheme. PowerInfer also skips only in the feed-forward network. To skip 54\% of a layer, it must
skip 80\% of the weights there. This requires the model to use the ReLU function as its activation function. It also
needs trained predictors. \sys{} spreads the sparsity over all seven
projections and runs models with the SiLU function as they are.

\subsection{Significance of key designs}
\label{sec:eval-gain}

\begin{figure}[t]
\includegraphics[width=\columnwidth]{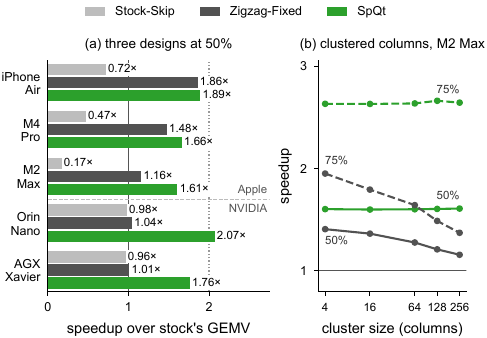}
\caption{The layout makes skipping pay; the list keeps it balanced. Speedup of a Llama-3-8B layer's weight products
(Q4\_K) over stock's GEMVs: (a) three designs at 50\% (dotted: 2$\times$, half the bytes); (b) as the active columns
cluster, on the M2 Max.}
\label{fig:ops}
\end{figure}

To isolate the layout and the kernel, we time the weight products of a Llama-3-8B layer alone, in Q4\_K at 50\%
sparsity, each device in its own band and the iPhone's GPU held at its medium performance state (\autoref{fig:ops}a). \sys{}'s kernel is given its active-column lists.

\paragraph{The zigzag layout} Skipping pays only on the zigzag layout. On the original
layout, \emph{Stock-Skip} is slower than stock on every device: it runs at 0.17--0.72$\times$ of stock's speed on the
Apple GPUs and at 0.96--0.98$\times$ on the NVIDIA GPUs. This is because almost every block still holds nonzero
activations, so almost every byte is read (\autoref{tab:skip}), and the branch per activation adds work even where
nothing is skipped (\S\ref{sec:motiv-conflict}). On the zigzag layout, \emph{Zigzag-Fixed} runs at 1.01--1.86$\times$ of
stock's speed and \sys{} at 1.61--2.07$\times$.

\paragraph{The zigzag sparse GEMV kernel} With the same band setting, dividing the active-column list evenly gains
2--39\% over \emph{Zigzag-Fixed} on the Apple GPUs and 74--99\% on the NVIDIA GPUs. This is because \emph{Zigzag-Fixed}
tests every activation, and the lanes of a zeroed activation idle; with the list, every lane works on an active column.

We evaluate load balancing on the active columns in a controlled test (\autoref{fig:ops}b). We keep the number of
active columns fixed, move them into clusters of 4 to 256 neighboring columns, and time the layer with \sys{} and with
\emph{Zigzag-Fixed} on the M2 Max. As the clusters grow, the fixed ranges of \emph{Zigzag-Fixed} fill unevenly, and its
speedup falls from 1.41 to 1.16$\times$ at 50\% and from 1.95 to 1.37$\times$ at 75\%. \sys{} stays at 1.6$\times$ at
50\% and 2.6$\times$ at 75\% for every cluster size, since each threadgroup still takes an equal slice of the list.
Real activations cluster less than this: they are spread almost at random, except in the input of $W_o$
(\S\ref{sec:motiv-conflict}).

\begin{figure}[t]
\includegraphics[width=\columnwidth]{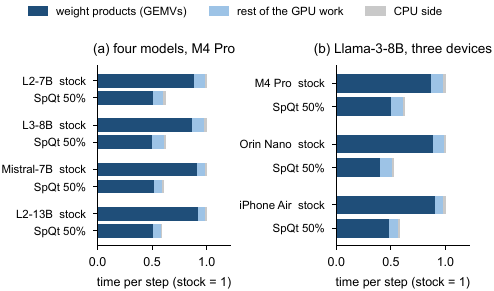}
\caption{Sparsity shrinks only the weight products, and the decode step keeps the time they save: stock against 50\%
in Q4\_K for (a) four models on the M4 Pro and (b) Llama-3-8B on three devices.}
\label{fig:breakdown}
\end{figure}

\paragraph{Fused sparsity detection} Building the active-column lists adds little
time to the decode step. \autoref{fig:breakdown} splits a step into the weight products, the rest of the GPU work and
the CPU side. At 50\%, the time that the whole step saves is 98--102\% of the time that the weight products save, on
the M4 Pro, the Orin Nano and the iPhone Air. For example, on the M4 Pro the weight products shrink from 17.7 to
10.1\,ms and the step from 20.4 to 12.8\,ms. The rest of the GPU work, which now includes building the lists, stays
within 1\,ms of stock's. This is because the lists are built inside the normalization, SiLU and residual kernels that
stock also runs. The work outside the weight products caps the gain: attention, the output layer and the normalizations
take 1.4--2.4\,ms per step on the M4 Pro, 5.9\,ms on the iPhone Air and 13\,ms on the Orin Nano.

\subsection{Sensitivity analysis}
\label{sec:eval-sens}

\begin{figure}[t]
\includegraphics[width=\columnwidth]{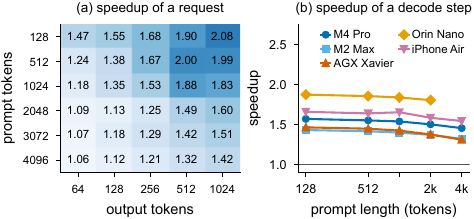}
\caption{A request gains less when its prompt is long, and a decode step gains slightly less as the context grows.
Llama-3-8B (Q4\_K) at 50\%: (a) speedup of whole requests on the iPhone Air; (b) speedup of a decode step after prompts
of 128 to 4,096 tokens.}
\label{fig:context}
\end{figure}

\paragraph{Prompt and output length}\label{sec:eval-requests} A whole request gains less than its decode steps, because
\sys{} processes a prompt densely. On the iPhone Air (\autoref{fig:context}a), a request with a
128-token prompt and 64 output tokens runs 1.47$\times$ faster at 50\%, while one with a 4,096-token prompt gains only
1.06$\times$. The same two requests gain 1.44 and 1.00$\times$ on the M4 Pro and 1.35 and 1.05$\times$ on the M2 Max.
On the phone, requests with long outputs gain more than a decode step measured alone (up to 2.08$\times$ against
1.66$\times$), because the phone throttles during a long output and \sys{}, doing less work per token, throttles later
than stock.

\paragraph{Context length} The decode speedup declines slowly as the context grows (\autoref{fig:context}b), because
attention grows with the context and sparsity does not reduce it. From a 128-token prompt to a 4,096-token one, the
speedup at 50\% falls by 7--11\% on the four devices that run both, for example from 1.57 to 1.46$\times$ on the M4 Pro.

\begin{figure}[t]
\includegraphics[width=\columnwidth]{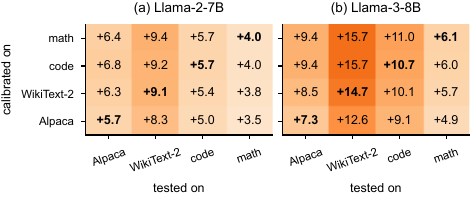}
\caption{Thresholds calibrated on one text hold on the others. Each cell is the perplexity increase (\%) over 0\% sparsity when
thresholds calibrated for 50\% sparsity on one text (rows) run on another (columns), in Q4\_K. Bold: the same text.}
\label{fig:calibration}
\end{figure}

\paragraph{Calibration text} Thresholds are calibrated once, offline, so we test whether they transfer across domains
(\autoref{fig:calibration}). We calibrate Llama-2-7B and Llama-3-8B in Q4\_K on four texts (Alpaca, WikiText-2, code
from MBPP~\cite{austin2021program} and math from GSM8K) and test each set of thresholds on all four.
Thresholds calibrated for 50\% sparsity on one text reach 47.6--51.2\% sparsity on every test text, for both models.
At the same target, calibrating on the test domain brings no gain: Alpaca's
thresholds, TEAL's default, reach slightly less sparsity (47.6--49.7\%) and cost the least perplexity on every test
text.
For the same test text, the cost differs across the four calibration texts by at most 1.1 points for Llama-2-7B and by
at most 3.1 points for Llama-3-8B, so one calibration set serves every domain.

\begin{figure}[t]
\includegraphics[width=\columnwidth]{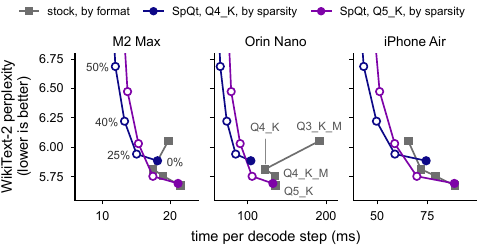}
\caption{Sparsity gives a better trade-off between accuracy and speed than a lower-precision format: Llama-3-8B's
perplexity against time per decode step, for stock in formats from Q5\_K to Q3\_K\_M, quantized with an importance
matrix, and for \sys{} in Q4\_K and Q5\_K at 0--50\% sparsity.}
\label{fig:frontier}
\end{figure}

\paragraph{Format and sparsity} Sparsity gives a better trade-off between accuracy and speed than a lower-precision
format (\autoref{fig:frontier}). The figure plots Llama-3-8B's perplexity against its time per decode step: stock in
formats of decreasing precision, from Q5\_K to Q3\_K\_M, and \sys{} in Q4\_K and Q5\_K at 0--50\% sparsity. Q4\_K\_M
and Q3\_K\_M are presets of llama.cpp's quantizer that mix formats within a model.
For this comparison, stock's files are quantized with llama.cpp's importance matrix, which shifts precision
within a block toward the columns with large activations and lowers stock's perplexity by 0.3--2.1\%. It has no
effect on the zigzag layout, where a block holds one column.
Q5\_K at 25\% sparsity is as accurate as the popular Q4\_K\_M and
takes 8--22\% less time per token across the five devices. It is also more accurate than
stock's Q4\_K and at least as fast.
Q4\_K at 25\% sparsity is more accurate than the lowest-precision format, Q3\_K\_M, and takes 10--56\% less time.
\section{Related work}
\label{sec:related}

\paragraph{Activation and neuron sparsity} For each token, a layer needs only the part of its weights that the
token's nonzero activations use. \emph{Predictor-based} systems learn which feed-forward neurons a token will use. DejaVu
skips the other neurons, and predicts attention heads as well~\cite{liu2023dejavu}, which Polar Sparsity extends to
batched serving~\cite{shrestha2025polarsparsityhighthroughput}; PowerInfer keeps the frequently
used neurons on the GPU and the rest on the CPU~\cite{song2024powerinfer}, and PowerInfer-2 brings the idea to
smartphones~\cite{xue2024powerinfer2}; LLM in a flash loads only the predicted neurons from flash
storage~\cite{alizadeh2024flash}, and Neuralink places neurons that fire together next to each other
there~\cite{wang2025neuralink}. These systems are sparsest on ReLU models, to which modern models are converted by
fine-tuning~\cite{mirzadeh2024relu, song2024prosparse, song2024turbosparse}; Q-Sparse trains models to be sparsely
activated~\cite{wang2024qsparse}. \emph{Training-free} methods threshold the activations of SiLU models by magnitude:
CATS in the feed-forward network~\cite{lee2024cats}, TEAL in every projection~\cite{liu2025teal}, and R-Sparse with a
low-rank correction~\cite{zhang2025rsparse}. Both groups decide \emph{which} activations a projection may skip. The
kernels of the training-free methods compute on 16-bit weights; PowerInfer runs a 4-bit format with a fixed range of
neurons per thread block, and \S\ref{sec:eval-powerinfer} compares it with \sys{}. Similar to TEAL, \sys{} thresholds
the input of every projection and needs no training. Unlike these works, it makes activation sparsity practical on
quantized weights: it stores every projection in the zigzag layout, in the existing formats, and keeps the GPU's load
balanced.

\paragraph{Quantization} Weight quantization made on-device LLMs practical. LLM.int8()~\cite{Dettmers2022llmint8},
GPTQ~\cite{frantar2022gptq}, AWQ~\cite{lin2024awq}, SmoothQuant~\cite{xiao2023smoothquant} and llama.cpp's
K-quants~\cite{Kawrakow2023kquant} choose the low-bit values; dense kernels such as Marlin~\cite{frantar2025marlin}
and serving systems such as QServe~\cite{lin2024qserve} run them fast; SpQR stores the weights that are hard to
quantize as sparse outliers~\cite{dettmers2024spqr}. SharQ also quantizes the activations, to 4-bit floating point: it
computes the largest in a sparse product and the rest in a dense one~\cite{meng2026sharq}. Group-wise formats such as
GPTQ's, AWQ's and the K-quants share
a scale among consecutive weights of one row, which a dot product with the input vector reads together. \sys{} keeps
llama.cpp's formats. Unlike these works, its zigzag layout quantizes the weights along the columns and stores each
column's blocks together, so that a sparse GEMV reads only the active columns.

\paragraph{Weight sparsity and sparse kernels} Weight sparsity is \emph{static}: pruning removes weights from a model
offline, either unstructured (SparseGPT~\cite{frantar2023sparsegpt}, Wanda~\cite{sun2024wanda}) or in N:M patterns
that sparse tensor cores accelerate~\cite{mishra2021accelerating}. GPU kernels for unstructured weight
sparsity~\cite{xia2023flashllm, fan2025spinfer} and sparse tensor compilers and
formats~\cite{kjolstad2017taco, ye2023sparsetir} handle sparsity that is known before the computation. Unlike these works, \sys{} handles \emph{dynamic} sparsity: its weight matrices stay dense, and the active
columns change with every token.

\paragraph{On-device LLM systems} Inference engines such as llama.cpp~\cite{ggerganov2022llamacpp}, MLC
LLM~\cite{mlc-llm} and MNN-LLM~\cite{wang2024mnnllm} run quantized models densely on the CPUs and GPUs of personal
devices. A measurement study on phones and Jetson boards finds that token generation is bound by memory bandwidth and
that sustained use is limited by energy and heat~\cite{laskaridis2024melting}.
Recent systems use a device's other processors: llm.npu offloads
prompt processing to the NPU~\cite{xu2025llmnpu}, and HeteroInfer runs the GPU and the NPU
together~\cite{chen2025heteroinfer}. STI pipelines the loading of a model with its execution under a memory
budget~\cite{guo2023sti}. These systems choose where and when a model's weights are computed. \sys{} is built into
llama.cpp and runs on the GPU alone, where it reduces the weights that a decode step reads.
\section{Conclusions}
\label{sec:conclusion}

We present \sys{}, a novel system that makes activation sparsity practical for quantized LLMs on personal devices.
\sys{} contributes three techniques. The zigzag layout cuts each weight matrix into bands of rows; within each
band, it quantizes the weights along the columns and stores each column's blocks together, in the existing formats.
The zigzag sparse GEMV kernel computes on this layout and keeps the GPU's load balanced. Fused sparsity detection finds the nonzero
activations inside kernels that each decode step already runs. With them,
\sys{} decodes 1.45--2.15$\times$ faster than llama.cpp at 50\% sparsity on five Apple and NVIDIA devices, with up to
2.6$\times$ less energy per token. A device can thus choose a sparsity level in addition to a quantization format,
trading a small loss in accuracy for speed and energy.

\begin{acks}
In accordance with arXiv's policy on generative AI language tools, we disclose the use
of an LLM in writing this paper. The LLM drafted and revised text in all sections, working from our outlines, notes
and measurement results. We directed the writing, reviewed and edited the text line by line, and take full
responsibility for the content.
\end{acks}

\end{document}